\documentclass[prl,twocolumn,floatfix,showpacs,superscriptaddress]{revtex4-1}

\usepackage[latin1]{inputenc}
\usepackage{amsmath}
\usepackage{mathrsfs}

\usepackage[usenames,dvipsnames]{color}
\usepackage{amsfonts}

\usepackage{amssymb,latexsym} 

\usepackage{enumerate}

\usepackage{graphicx}

\usepackage{hyperref}
\usepackage{multirow}

\begin{document}
\title{Percolation on trees as a Brownian excursion:\\from Gaussian to
Kolmogorov-Smirnov to Exponential statistics }

\author{Francesc Font-Clos}
\affiliation{ISI Foundation,
Via Alassio 11/c,
10126 Torino, Italy
}

\author{Nicholas R.~Moloney}
\affiliation{London Mathematical Laboratory,
14 Buckingham Street,
London WC2N 6DF, UK
} 

\begin{abstract}
We calculate the distribution of the size of the percolating cluster
on a tree in the subcritical, critical and supercritical phase. We do
this by exploiting a mapping between continuum trees and Brownian
excursions, and arrive at a diffusion equation with suitable boundary
conditions. The exact solution to this equation can be conveniently
represented as a characteristic function, from which the following
distributions are clearly visible: Gaussian (subcritical),
Kolmogorov-Smirnov (critical) and exponential (supercritical). In this
way we provide an intuitive explanation for the result reported in
R. Botet and M. P{\l}oszajczak, Phys.~Rev.~Lett \textbf{95}, 185702
(2005) for critical percolation.
\end{abstract}

\date{\today}

\maketitle

\section{Introduction}

Mappings that connect seemingly unrelated models are regularly used
in statistical physics. They not only facilitate calculations but
provide additional intuition about the original models. Well-known
examples include the Ising model interpretation for a lattice
gas\cite{Cardy:1996}, the bosonic interpretation for the partitioning
of an integer into summands\cite{AuluckKhotari:1946}, or the Coulomb
gas interpretation for the eigenvalues of certain random
matrices\cite{Mehta:2004}. In practice, the coincidence in
distribution of an observable in one model may spur the search for a
mapping to its counterpart observable in another model. In this
Letter, we adopt this route inspired by the Kolmogorov-Smirnov (KS)
distribution, which has been noticed in a variety of contexts. It is
the distribution of a test statistic for comparing between empirical
and theoretical distribution functions, and is well understood to
describe the absolute maximum value of a Brownian
bridge\cite{Doob:1946}. With this insight, it can be related to other
Brownian observables\cite{BianePitmanYor:2001}. More surprisingly, it
is the distribution of the integrated mean-squared fluctuations of a
periodic Brownian signal\cite{Watson:1961}, and therefore also
accounts for the roughness of a 1d periodic Edwards-Wilkinson
interface in the steady state\cite{FoltinETAL:1994}. It also describes
the sizes of clusters in a mean-field aggregation
process\cite{BotetPloszajczak:2005}.

This paper focuses on the finite-size scaled distribution of
percolating cluster sizes on a Bethe lattice at the critical point,
first computed by Botet and
P{\l}oszajczak\cite{BotetPloszajczak:2005,Botet:2011} to be the KS
distribution. We demonstrate that there is indeed a connection to
Brownian motions, thereby obtaining a more intuitive understanding of
their result. In this way, we demystify the coincidence in
distribution and bring the result into the existing fold of knowledge
about Brownian motions and associated
observables\cite{BianePitmanYor:2001,Majumdar:2005}.
 
\section{Setting up the problem}

We consider site percolation on a finite Bethe lattice of size $L$,
coordination number $z$ and site occupation probability $p$, with
critical occupation probability $p_c
=1/(z-1)$\cite{StaufferAharony:1994}. There is a distinguishable site
at the center, called the root, from which distances or `heights' to
other sites can be measured. With a subsequent mapping to Brownian
excursions in mind, it is convenient to define the root to be at
$h=1$. Neighboring sites in the first generation are then at $h=2$,
sites at the boundary at $h=L$, etc. We say that the system
{percolates} if there is at least one path of occupied sites from the
root to any boundary site. The percolating cluster, containing the
root, can be thought of as a rooted tree of fixed height.

The size $S$ of the percolating cluster is the number of sites
forming the cluster that contains the root (which may include more
than one path to the boundary).  If the system does not percolate, we
set $S=0$ for convenience. 

In what follows, we will be concerned with the distribution of the
size of the percolating cluster \emph{given that} the system
percolates,
\begin{equation}
\mathcal{P}(s) \equiv \mathrm{Prob}[S = s : S>0].
\end{equation}
In particular, we will direct our attention to the limit of large
system sizes $L\gg1$. A well-normalized limiting distribution for the
rescaled percolating cluster size will therefore depend on how $S$
scales with $L$ for different values of $p$.

\section{Mapping to Brownian Excursion}

The well-known \emph{depth-first search}\cite{Aldous1993} (also known
as a Harris walk) gives a bijection between rooted trees and
excursions. Here we make use of an asymptotic version of this
technique for $L \gg 1$ to map the percolation problem to a Brownian
excursion with specific boundary conditions.

\begin{figure}
\includegraphics[width=\columnwidth]{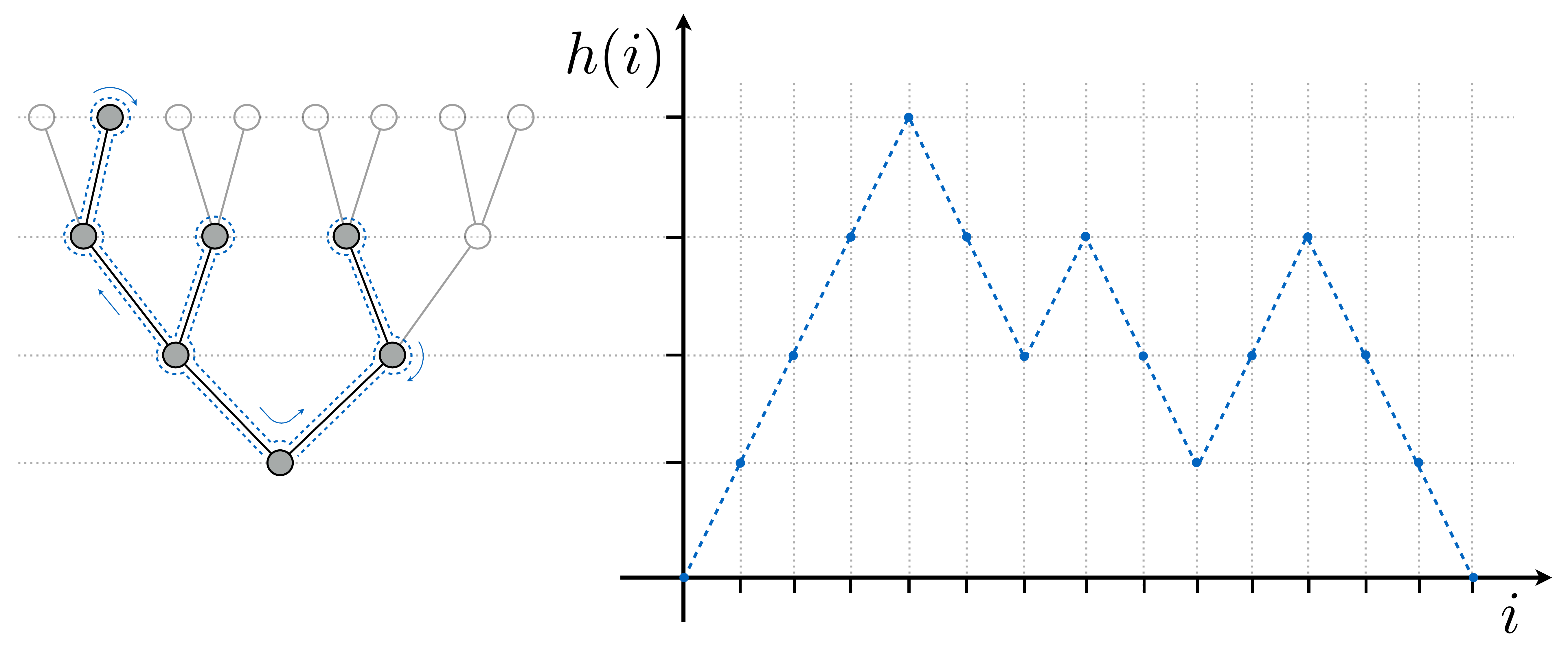}
 \caption{The correspondence between trees and positive paths in a
   system of size $L=4$. The percolating cluster has size $S=7$,
   and the associated walk has duration $T=2S=14$.  Left: The
   percolating cluster, traversed in depth-first search order, as
   indicated by the arrows next to the edges.  Right: The
   associated path, with an initial and final step appended so that
   the walk does not return to $h=0$ before the entire tree has been
   traversed.}
 \label{F:mapping}
 \end{figure}

Consider a realization of the percolating cluster, traversed in
depth-first search order starting from the root at time $i=0$, see
Fig.~\ref{F:mapping}. Each \emph{edge} is traversed exactly twice,
once in each direction, and the pairs $(i,h(i))$ form a positive walk
of length $2S$. To have a well-defined excursion that terminates at
$h=0$ once the tree has been traversed and not before, an up-step and
a down-step is appended to the beginning and end of the path.
For large $L$, the walk $(i,h(i))$ can be approximated by a Brownian
excursion. If we further think of percolation on the Bethe lattice as
a Galton-Watson branching process with binomial offspring distribution
$X\sim\mathcal{B}(z-1,p)$, then the drift $v$, diffusion constant $D$
and initial condition $x_0$ of the associated excursion can be
determined simply by inspecting and matching some well-known
results. For example, the probability for an unbiased excursion to
reach level $k$ before 0, starting from $x_0$, is $x_0/k$ (the
gambler's ruin probability); while the probability for a critical
branching process to survive at least $k$ generations is
$2/(\mathrm{Var}[X] k)$\cite{Harris1963,AthreyaNey2004}. This result,
together with the matching of the first two cumulants of the total
population of a subcritical branching process\cite{SuhovKelbert2014}
with those of the corresponding Brownian excursion, leads to the
identifications:
\begin{equation}
v  = \frac{\mathrm{E}[X]-1}{\mathrm{Var}[X]},\quad
D  = \frac{1}{\mathrm{Var}[X]},\quad
x_0  = \frac{2}{\mathrm{Var}[X]}, 
\label{E:lockdown}
\end{equation}
where $\mathrm{E}[X]=(z-1)p$ and $\mathrm{Var}[X]=(z-1)p(1-p)$. This
is in agreement with the known rescaling of the so-called contour
process~\cite{Bennies2000}.  In addition, the system is conditioned to percolate,
so that the associated walk must reach height $h(j)=L$ for some
$0<j<2S$ before returning to the origin for the first time at $i=2S$.
In summary, the large $L$ limit of the walk $(i,h(i))$ corresponds to
a Brownian excursion of maximum height $L$, drift $v$ and diffusion
constant $D$.  The (random) duration $T$ of this Brownian excursion
equals $2S$ by construction.

Proper convergence of a positive walk to a Brownian excursion is
formally only reached after rescaling lengths appropriately, but we
will only take this step at the very end of our calculations, once the
scaling of $S$ with $L$, as a function of $p$, has been calculated.
Formal proofs of convergence in similar setups can be found in the
related literature on \emph{continuum random trees}. For instance,
Aldous\cite{Aldous1993} shows that if certain families of trees are
constrained to have a fixed large number of nodes, then its associated
Harris walks converge to the standard Brownian excursion of length 1;
while Le Gall\cite{LeGall:2005} also considers constraints related to
the maximum height.  In both cases it is shown that, under certain
mild conditions, all families of trees (i.e., all lattices) lead to
the same universal results.

We have thus set up a mapping between percolation in a finite Bethe
lattice and a Brownian excursion with a reflecting boundary, and
explicitly related their parameters, Eq.~\eqref{E:lockdown}. In what
follows, we will compute the distribution of $T$ in the Brownian
excursion setting, and show that the results agree, as expected, with
simulations of percolation in a Bethe lattice.

\section{Brownian Excursion}

\begin{figure}[htbp]
\begin{center}
\includegraphics[width= \columnwidth]{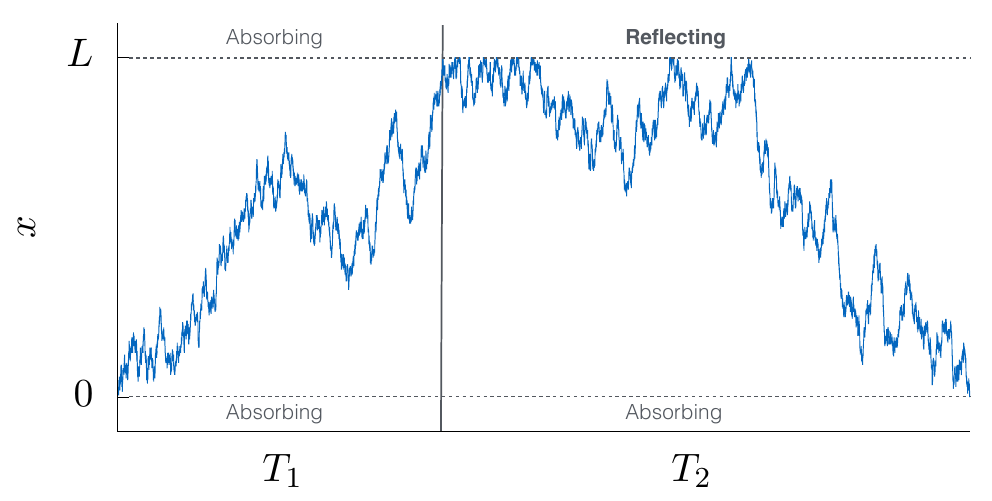}
\end{center}
\caption{A Brownian excursion, conditioned to reach $x(t) = L$ for
  some $0<t<T=T_1+T_2$, decomposed into (i) the path from $x=0$ that
  reaches $x=L$ for the first time, (ii) the return path from $x=L$
  back to $x=0$. 
  \label{F:path_decomposition}}
\end{figure}

The Brownian excursion of fixed height $L$ corresponding to our
problem can be decomposed into two paths (see
Fig.~\ref{F:path_decomposition}): (i) a first path from the origin
$x(0)=0$ at time $t=0$ to the boundary $x(T_1)=L$, taking time $T_1$,
and (ii) a return path from the reflecting boundary $x(T_1)=L$ to the
origin $x(T)=0$, taking time $T_2=T-T_1$. Note that the first path
involves a conditional exit at $x=L$ such that that boundary is
effectively absorbing for $t<T_1$. Readers experienced with Brownian
processes will recognize $T_1$ as the hitting time to reach $L$ of a
3d Bessel process\cite{BianePitmanYor:2001}.

\subsection{Diffusion equation with drift}

The distributions of $T_1$ and $T_2$ are most conveniently calculated
by working with the Laplace transform of the diffusion equation with
drift:
\begin{equation}
\left[
-D\partial_{xx} + v \partial_{x} + \lambda 
\right] \hat \phi(x,\lambda) = \delta(x-x_0) 
\end{equation}
where $\phi(x,\lambda)$ is the Laplace transform with respect to time
of the density function with initial condition $x(0)=x_0$. According
to the usual recipe\cite{Redner:2001}, this equation is solved with
combinations of exponentials respecting the boundary conditions, and
fluxes are computed at the relevant boundaries. These calculations
furnish the Laplace transformed first passage densities $j_1(\lambda)$
and $j_2(\lambda)$ associated with $T_1$ and $T_2$.

\subsection{First path: A + A}

To calculate $T_1$ for the first path of the decomposition, absorbing
boundaries are required at both at $x=0$ and $x=L$,
\begin{align}
\hat \phi_{AA}(0,\lambda) &= 0 \\
\hat \phi_{AA}(L,\lambda) &= 0, 
\end{align}
together with a conditional exit $x=L$. $\hat\phi_{AA}$ satisfying the
absorbing boundaries takes the form
\begin{equation}
\hat \phi_{AA} (x,\lambda) =
\frac{2 e^{v  (x-x_0)/2D}}{\rho \sinh(\rho L/2D)}
   \sinh(\rho x_0/2D) \sinh(\rho(L-x)/2D)
\label{E:T1}
\end{equation}
where $\rho := \sqrt{v^2+4\lambda D}$. The conditional exit probability
at $x=L$ is\cite{Redner:2001}
\begin{equation}
  \mathrm{Prob}(\text{exit at } x = L) = \frac{1-e^{-vx_0/D}}{1-e^{-Lv/D}}
\end{equation}
so that the flux at the boundary $x=L$ is
\begin{align}
  j_1(\lambda) &= \lim_{x_0\to 0} \frac{-D
    \hat\phi'_{AA}(L,\lambda)}{\mathrm{Prob}(\text{exit at } x = L)} \\
&= \frac{\rho}{v} \frac{\sinh (Lv/2D)} {\sinh
    (\rho L/2D)}
\end{align}

\subsection{Return path: A + R}

To calculate $T_2$ for the second path of the decomposition, an
absorbing boundary is required at $x=0$
\begin{equation}
\hat \phi_{AR}(0,\lambda) = 0
\end{equation}
and a Dirac $\delta$-pulse is injected at $x = L$ as an initial
condition, which in Laplace space is equivalent to the boundary
condition
\begin{equation}
  -v \hat\phi_{AR}(L,\lambda) + D \hat\phi'_{AR}(L,\lambda) = 1
\end{equation}
$\hat\phi_{AR}$ satisfying these boundary conditions is
\begin{equation}
  \hat\phi_{AR}(x,\lambda) = \frac{2e^{v(x-L)/2D} \, \sinh(\rho x/2D)}
  {\rho \cosh (\rho L/2D) - v \sinh (\rho L/2D)},
\end{equation}
so that the flux at the boundary $x=0$ is
\begin{align}
  j_2(\lambda) &= D \hat\phi'_{AR}(0,\lambda) \\
  &= \frac{\rho \, e^{-Lv/2D}}{\rho \cosh (\rho L/2D) - v \sinh (\rho L/2D)}.
\end{align}

\subsection{Entire excursion}

The total time to go from the origin to the boundary at $x=L$, and
then from the boundary back to the origin is $T=T_1+T_2$. Since $T_1$
and $T_2$ are independent, the Laplace transform of the density of
the total time $T$ is the product
\begin{align}
  j(\lambda) &= j_1(\lambda) \cdot j_2(\lambda) \notag \\
  &= \frac{\rho^2}{v} \frac{\sinh (Lv/2D)} {\sinh (\rho L/2D)}
  \times \notag \\
  &\frac{e^{-Lv/2D}}{\rho \cosh (\rho L/2D) - v
    \sinh (\rho L/2D)} \label{E:full_solution}
\end{align}

\subsection{Critical case}

Critical percolation with $p=p_c = 1/(z-1)$ maps to Brownian motion
with zero drift. Taking the limit $v\to0$ in
Eq.\eqref{E:full_solution},
\begin{align}
  j_1^{(c)}(\lambda) &= \frac{L \sqrt{\lambda/D}}{\sinh(L\sqrt{\lambda/D})} \\
  j_2^{(c)}(\lambda) &= \frac{1}{\cosh(L\sqrt{\lambda/D})} \\
  j^{(c)} (\lambda) &= \frac{2L \sqrt{\lambda/D}}{\sinh(2L\sqrt{\lambda/D})}.
\end{align}
Interestingly, $j_1^{(c)}$ and $j^{(c)}$ are Laplace transforms of the
same KS distribution on different scales. $j_2^{(c)}$ is also the
Laplace transform of many observables of Brownian motions,
see\cite{BianePitmanYor:2001} for a review.

\subsection{General case}

\begin{figure}
\begin{center}
\includegraphics[width=\columnwidth]{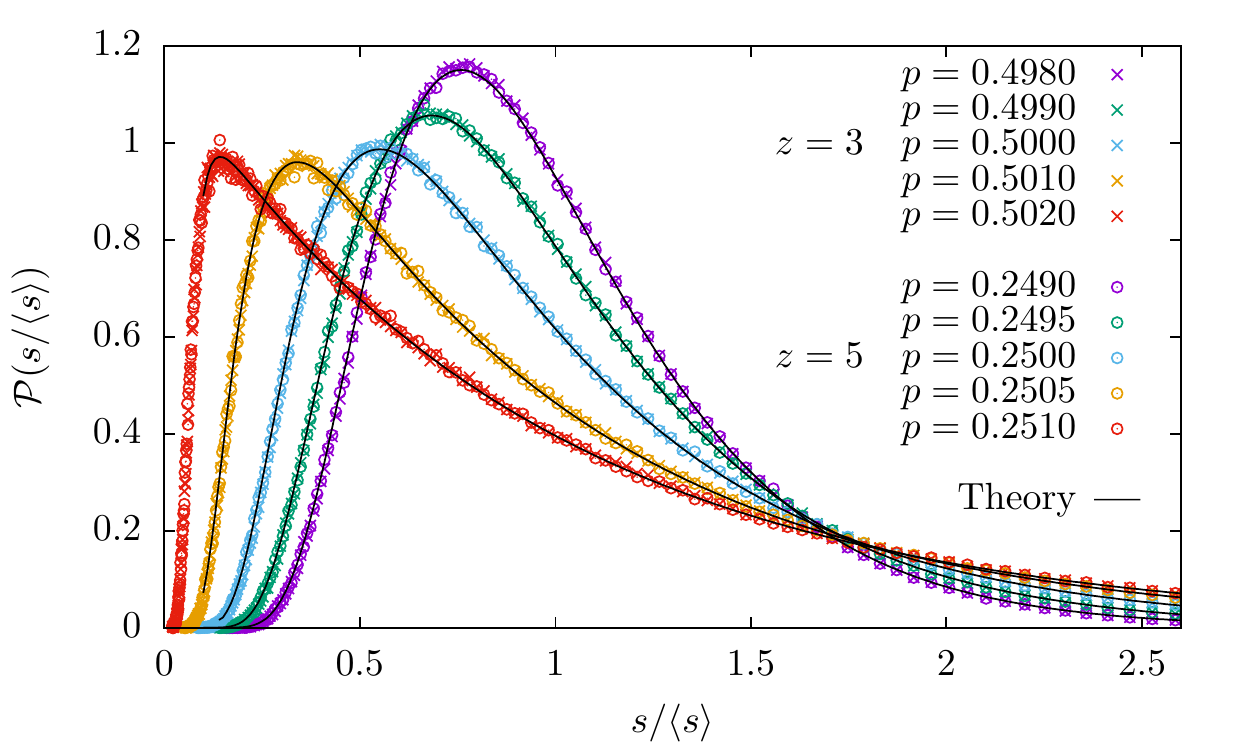}
\end{center}
\caption{The rescaled distribution of the percolating cluster size,
  for different values of $z$ and $p$ and fixed system size
  $L=10^3$. Simulations (colored symbols) agree with theory (black
  line), which is obtained via numerical inversion of
  Eq.\eqref{E:full_solution}.~\label{F:numerical_inversion}}
\end{figure}

For general choices of $v$ (or, equivalently, $p$), the full solution
in Eq.~\eqref{E:full_solution} cannot be inverted explicitly, and one
must resort to numerical inversion. Fig~\ref{F:numerical_inversion}
shows a family of percolating cluster size distributions, all rescaled
by their means.
The distributions are independent of coordination number $z$ as
expected, but clearly depend on $p$. For $p\neq p_c$ (and fixed $L$),
the distributions can be thought of as flowing away from the
non-trivial fixed point, represented by the KS distribution (middle,
blue curve).

To better understand these flows towards trivial fixed points
(in the language of renormalization-group (RG)) for $p\to 0$ and $p
\to 1$, it is convenient to standardize the distributions to zero mean
and unit standard deviation. Since Eq.~\eqref{E:full_solution} is the
moment generating function for $T$, these cumulants can be constructed as
usual by differentiating once (first moment) or twice (second moment)
with respect to $\lambda$, and taking $\lambda \to 0$. For the mean
$\mu := \langle T \rangle$ we find
\begin{equation}
  \mu \, \frac{v^2}{D} = \mathcal{F}(Lv/D),
\end{equation}
where the scaling function
\begin{equation}
  \mathcal{F}(x) = e^x - 3 + \frac{2x}{e^x-1}
\end{equation}
behaves as $e^x$ for $x\gg 1$, $x^2$ for $x \to 0$, and $x$ for $x\ll
-1$. For fixed $v$ and $D$, the scaling behavior of $\mu$ is therefore
\begin{equation}
  \mu = \begin{cases} 
\frac{D}{v^2} e^{Lv/D}, &\quad L \gg 1, v>0 \\
\frac{2}{3D} L^2, &\quad L \gg 1, v=0 \\
\frac{2}{|v|} L, &\quad L \gg 1, v < 0.
\end{cases}
\end{equation}
Expressions can likewise be obtained for the standard deviation
$\sigma := \sqrt{\langle T^2 \rangle - \langle T \rangle^2}$ (not
shown).

Given $\mu$ and $\sigma$, we arrive at a standardized characteristic
function $\mathcal{G}(k;Lv/D)$ from Eq.~\eqref{E:full_solution} by (i)
replacing $\lambda \to -ik$, to change from Laplace to Fourier space,
(ii) multiplying by $\exp(-ik\mu)$ to translate the distribution to
zero mean, and (iii) replacing $k \to k/\sigma$, to scale the
distribution to unit standard deviation. Note that the ratio
$\mu/\sigma$ is a function of $Lv/D$ alone. Thus, $\mathcal{G}(k;Lv/D)$
describes a family of probability distributions parameterized by $Lv/D$.

\begin{figure}
\begin{center}
\includegraphics[width=\columnwidth]{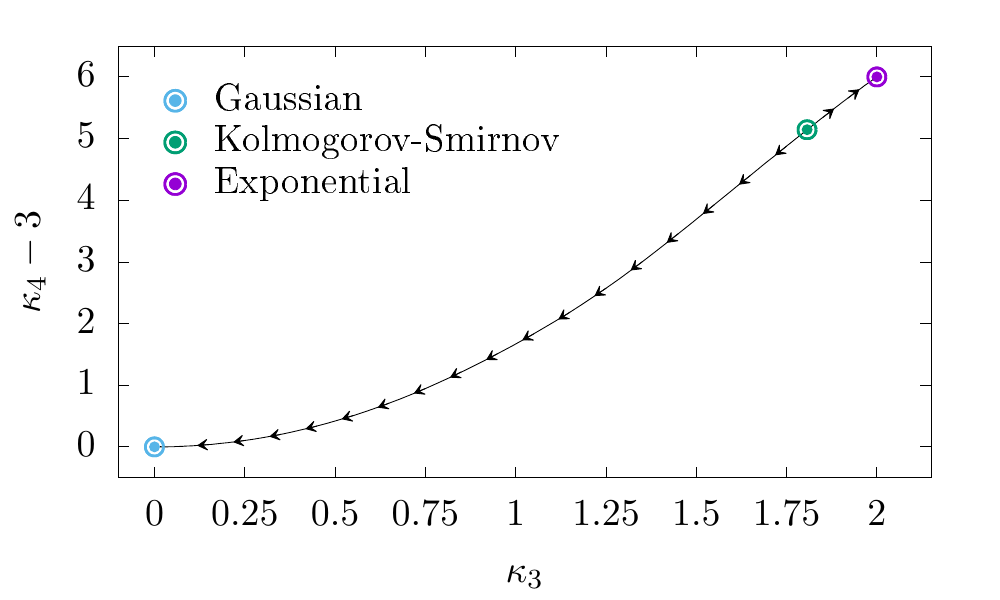}
\end{center}
\caption{Centered skewness $\kappa_3$ and kurtosis $\kappa_4-3$ for
  the standardized solution described by $\mathcal{G}(k;Lv)$. Solid
  circles from left to right correspond to the fixed points: Gaussian
  (subcritical), KS (critical), Exponential
  (supercritical). \label{F:skewness_kurtosis}}
\end{figure}

As a result of this standardization, all distributions
$\mathcal{G}(k;Lv/D)$ have zero mean and unit standard
deviation. However, higher cumulants $\kappa_n$ will depend on
$Lv/D$. These differences can be visualized by plotting the skewness
$\kappa_3$ versus the kurtosis $\kappa_4-3$, as depicted in
Fig.~\ref{F:skewness_kurtosis}. Three regimes are particularly
noteworthy: (i) $Lv/D \to -\infty$ (subcritical fixed point), (ii)
$Lv/D \to 0$ (critical fixed point), (iii) $Lv/D \to \infty$
(supercritical fixed point). With the help of Mathematica, we can
obtain full expressions for $(\kappa_3,\kappa_4-3)$ as a function of
$Lv/D$. In the three above regimes these reduce to (i) $(2,6)$, (ii)
$(4\sqrt{10}/7,36/7)$, (iii) $(0,0)$, corresponding to exponential, KS
and Gaussian distributions, respectively. These three fixed points are
marked in Fig.~\ref{F:skewness_kurtosis} as solid circles.
A heuristic explanation for the sub and supercritical limits is as
follows: the overall behavior of $T$ is dominated by $T_2$ ($T_1$ is
insensitive to the sign of $v$, which can be seen from the fact that
\eqref{E:T1} is an even function of $v$). When $v<0$, $T_2$ is
approximately the ballistic travel time from $x=L$ to $x=0$, with a
Gaussian correction coming from diffusion. When $v>0$, $T_2$ is
dominated by the rare event that the motion escapes the drift that
keeps returning it to the reflecting boundary at $x=L$. This rare
event process is consistent with exponential tails.

We briefly comment on how our method differs from that of Botet and
P{\l}oszajczak\cite{BotetPloszajczak:2005}, the full details of which
are presented in\cite{Botet:2011}. In their method, a recursion is set
up between successive generations of the Bethe lattice, encoding the
statistical weights of configurations conditioned to percolate. The
recursion relation is then Laplace transformed and rescaled by mean
percolating cluster size, to yield expressions which are then analyzed
asymptotically in the limit of large system size for the critical case
$p = p_c$\footnote{Off criticality, Botet reports (but does not show)
  Gaussian statistics, which we find only in the subcritical case}. In
contrast, our method makes use of a passage from continuum trees to
Brownian excursions at the starting point of the calculation, such
that the parameters of the Bethe lattice are reincorporated into the
diffusion constant and drift of the resulting motion. This makes the
subsequent analysis arguably more intuitive, and
Eq.\eqref{E:full_solution} gives the entire cluster size distribution
for all regimes from the solution of a diffusion problem with
drift. Our approach should be universal across different types of
trees, in the sense that the fixed point distributions (suitably
rescaled) do not depend on underlying microscopic details.

\section{Conclusion}

We have calculated the distribution of the size of the percolating
cluster on a tree by interpreting the problem as a type of Brownian
excursion. In this way, we give an intuitive explanation for the
coincidence in distribution (first noted
in\cite{BotetPloszajczak:2005}) with other observables associated with
Brownian motions\cite{BianePitmanYor:2001}. The analysis can be
extended off criticality by adding a drift term to the associated
diffusion equation. The resulting flows in the space of distributions
can be captured by tracking the skewness and kurtosis. Our exact
calculation makes an investigation of the various regimes possible in
full detail. We expect that such mappings can be used to investigate
further properties of branching processes.

\begin{acknowledgments}
The authors would like to thank \'{A}lvaro Corral, Gregory Schehr and
Zoe Budrikis for useful discussions.
\end{acknowledgments}

\bibliography{articles,books}

\end{document}